\documentclass[superscriptaddress,aps,prl,showkeys,twocolumn,10pt]{revtex4}
\usepackage{graphicx}
\usepackage[sort&compress]{natbib}
\usepackage{subfigure}
\usepackage{amsmath}
\usepackage{amssymb}
\usepackage{amsfonts}
\usepackage{rotating}
\usepackage{cancel}
\usepackage{lipsum}
\usepackage{mathtools}
\usepackage{color}
\usepackage{bbm}
\usepackage{dsfont}
\usepackage{bbold}
\usepackage{multirow}
\usepackage{ulem}

\begin{document}

\title{Reply to ``Comment on Sequential Single-pion Production Explaining the dibaryon $d^*(2380)$ peak''}
\date{\today}

\author{R. Molina}
\email{Raquel.Molina@ific.uv.es}
\affiliation{Departamento de F\'{\i}sica Te\'orica and IFIC,
Centro Mixto Universidad de Valencia-CSIC,
Institutos de Investigaci\'on de Paterna, Aptdo. 22085, 46071 Valencia, Spain}   
\author{Natsumi Ikeno}
\email{ikeno@tottori-u.ac.jp}
\affiliation{Departamento de F\'{\i}sica Te\'orica and IFIC,
Centro Mixto Universidad de Valencia-CSIC,
Institutos de Investigaci\'on de Paterna, Aptdo. 22085, 46071 Valencia, Spain}   
\affiliation{Department of Agricultural, Life and Environmental Sciences, Tottori University, Tottori 680-8551, Japan}
\author{Eulogio Oset}
\email{Eulogio.Oset@ific.uv.es}
\affiliation{Departamento de F\'{\i}sica Te\'orica and IFIC,
Centro Mixto Universidad de Valencia-CSIC,
Institutos de Investigaci\'on de Paterna, Aptdo. 22085, 46071 Valencia, Spain}   
\begin{abstract}
In a comment to our paper on ``Sequential single-pion production explaining the dibaryon $d^*(2380)$ peak'' the authors provide arguments that apparently invalidate our claims and present what they call ``proofs'' of the dibaryon explanation of the $np\to \pi^0\pi^0d$ fusion reaction. In this reply we refute the arguments of the comment and show that no existing experiment proves the dibaryon nature of the peak of the fusion reaction.
\end{abstract}
\maketitle

The study of the $np\to \pi^0\pi^0d$ and $np\to \pi^+\pi^-d$ reactions \cite{WASA-at-COSY:2012seb,WASA-at-COSY:2011bjg} showed an unexpected narrow peak in the cross section around $2380$ MeV, which has been attributed to a dibaryon by the experimental team, branded $d^*(2380)$. It is interesting to remark that based on earlier data of the reaction, the peak had been attributed to a reaction mechanism based on sequential one pion production, $np\to \pi^-pp\to \pi^-\pi^+d$, together with $np\to\pi^+nn\to\pi^+\pi^-d$, in the work of Bar-Nir et al. \cite{barnir}. The second step, $pp\to\pi^+d$, was the object of theoretical investigation in \cite{brack,green,weise}, and was shown to be driven by $\Delta(1232)$ excitation. A reformulation of the idea of these works was done by us in Ref.~\cite{tsikeno}, from a Feynman diagrammatic point of view, showing that the process developed a triangle singularity (TS) \cite{landau,bayarguo}. This finding is relevant to the present discussion because it is well known that a TS gives rise to an Argand plot like the one of an ordinary resonance, even if the origin is a kinematical singularity and not the presence of a genuine physical state \cite{compass,guofeng}.

The idea of Bar-Nir et al. was retaken in \cite{seq} and, making some reasonable approximations and using experimental data on $np(I=0)\to pp \pi^-$ and $pp\to \pi^+d$ reactions, a peak was obtained for the $pp\to\pi^+\pi^-d$ reaction in qualitative agreement with experiment in the position, width and strength. Even with the approximations done, and the qualitative agreement found, such agreement, together with the result of Bar-Nir et al., can hardly be an accident and offer an alternative explanation of the peak observed in the experiment.

In between, a comment on the work of Ref.~\cite{seq} has appeared in the web, criticizing the approach and providing apparent ``proofs'' of the dibaryon resonance \cite{comment}. 

At present we are working on the description of the first step of the sequential mechanism, the $pn(I=0)\to pp\pi^-$ reaction, and a reply to the comment was reserved for the presentation of that work with a more quantitative description of the two step process. Anticipating a delay in this project, we found convenient not to delay unnecessarily a reply to the comment that we present here.

The comment contains eight points that we reply here one by one.
\begin{itemize}
 \item[i)] In point 1) of the comment the authors complain about us enlarging the errors in the $pn (I=0) \to \pi^- pp$ reaction. The reason for that is that the cross section for this reaction is obtained using isospin symmetry with the relation 
\begin{equation}
\sigma_{np(I=0) \to pp \pi^-} = ( \sigma_{np \to pp \pi^-} - \sigma_{pp \to pp \pi^0}/2)\ .\label{eq:sig}
\end{equation}

But the problem is that there are huge cancellations in this formula and the result is ten times smaller than each individual term. What we did is to assume a $5$\% uncertainty in the terms from isospin violation and determine the errors in the results. These are then systematic uncertainties that we think should have been considered by the experimentalists but they did not. So, we did. In the high energy of the spectrum, where the cross section falls down and produces the shape of the cross section, the systematic errors are much bigger than the statistical ones, hence it does not matter how the statistical ones are summed to them. The size is given by the systematic errors. In any case, the cross sections that we obtain have nothing to do with these errors.
 \item[ii)]  In point 2) the authors of the comment make a point about we not getting a precise description of the data. With the approximations that we did in Ref.~\cite{seq} we cannot pretend to get that precise agreement. We already consider an accomplishment that in such a complicated reaction we could get qualitatively a peak for the reaction at the right energy,  with the right width and the right strength at a qualitative level. 
 \item[iii)] The argument of this point is weak. First let us note that in our work of Ref.~\cite{tsikeno} for the $pp\to \pi^+d$ reaction we proved the dominance of the $^1D_2$ as found experimentally in \cite{arndt,Oh:1997eq}. Second, the argument states that because in the $np(I=0)\to\pi^-pp$ reaction the invariant mass of $\pi^-p$ is big, then the one of $pp$ is small and only accommodates $L=0,L=1$ waves, not $D$-waves necessary for the overlap of the two-step mechanism.It is interesting to make this argument more quantitative. We have two situations where $M(pp)$ is easily evaluated. They correspond to the case of $M(\pi^-p)\vert_{\mathrm{min}}=m_p+m_{\pi^-}$ and $M(\pi^-p)\vert_{\mathrm{max}}=\sqrt{s}-m_p$. In the first case the $p$ and $\pi^-$ move together in opposite direction to the other proton. In the second case one proton is produced at rest. The $M(pp)$ is trivially evaluated in these two cases and for the energy $T_p=1200$~MeV of Fig.~6 of Ref.~\cite{WASA-at-COSY:2017igi} we find,
 \begin{itemize}
 \item[a)] $M(pp)$ (at $M(\pi^-p)\vert_{\mathrm{min}}$) $=2239.47$~MeV, with an excess energy of the two protons of $362.9$~MeV.
 \item[b)] $M(pp)$ (at $M(\pi^-p)\vert_{\mathrm{max}}$) $=1920.2$~MeV, with an excess energy of $43.65$~MeV.
 \item[c)] There is another situation which also allows for an easy evaluation. This is at the peak of the distribution around $M(\pi^-p)=1370$~MeV where the $M(\pi^-p)$ from either of the protons is about the same and enhances the contribution in this region. There we have,
 $$\hspace{1cm} 2\,M^2(\pi^-p)+M^2(pp)=s+2\,m^2_p+m^2_{\pi^-}\ ,$$
 from where we get $M(pp)\simeq 1951$ MeV corresponding to about $75$~MeV excess energy for the two protons. Assuming relative distances of the produced two protons of $r\simeq 2.13$~fm, the radius of the deuteron, corresponding to the range of pion exchange, we find that the angular momentum, $L\sim r\times p$, can reach up to $L=6$ in the case of a), $L=2$ in the case b), and $L=3$ in the most favorable case corresponding to case c).
 \end{itemize}
 
 In the comment of \cite{comment}, $L=2$ is already ruled out and, without any calculation the sequential pion production cross section is deemed very small, contradicting the conclusions of \cite{barnir}. Actually, with the dominance of the Roper excitation, as claimed in \cite{comment}, we could already observe that $S=0$ for the two protons is the dominant mode in $np(I=0)\to\pi^-pp$, as in the second step of \cite{tsikeno} for the $pp\to \pi^+d$ reaction, and several $L$ values are allowed. Work continues along these lines since Roper excitation is not the only ingredient of the $np(I=0)\to\pi^-pp$ reaction. 
  \item[iv)] Point 4) is illustrative.  In two independent papers, \cite{wilkin} and \cite{miguel}, it was shown that there was a relationship between the $pn (I=0) \to \pi^+ \pi^- d$ reaction and the one where the $np$ of the deuteron would become free states.  The existence of the peak of the  $pn (I=0) \to \pi^+ \pi^- d$ reaction had as a consequence a peak also in $pn  \to \pi^+ \pi^- n p$ and related reactions at the same energy. However, this was the case independently of which is the reason for the peak in the fusion reaction. This is a key point. Actually the authors of the comment,  in order to calculate the cross section of the open reactions, used the results of these references and added the contribution to the results of the standard model for these reactions that they also use from \cite{luis}. Yet, they see that as an evidence of a dibaryon, while it was clearly shown in \cite{wilkin,miguel} that the new contribution was necessary whichever be the reason for the fusion reaction.
  \item[v)] In point 5) the comment complains that we do not calculate differential distributions. This is true, but we did not need that to prove our points. At the qualitative level that we worked, we showed that the distribution had to peak at small invariant masses of the two pions because we had two contributions: $pn (I=0) \to \pi^- pp$ followed by $pp \to \pi^+ d$, together with $pn (I=0) \to \pi^+ nn$ followed by $nn \to \pi^- d$. We could prove that when the momenta of the two pions are equal, the two amplitudes are identical and sum, producing a Bose enhancement.  In this case, the invariant mass of the two pions has its smallest value. This is why the cross section peaks at low $\pi \pi$ invariant mass. 
  \item[vi)] The comment claims that the model of Ref.~\cite{seq} cannot explain the observed pole in $^3D_3-^3G_3$ $np$ partial waves. This statement is incorrect. The  $NN$, $I=0$, phase shifts will be affected by the peak in the  $pn (I=0) \to \pi^+ \pi^- d$ reaction because one can have $pn(I=0) \to \pi^+ \pi^- d \to  pn(I=0)$, where $\pi^+ \pi^- d$ is in an intermediate state and contributes to the inelasticities. This will be particularly the case in the quantum numbers preferred by the $pn (I=0) \to \pi^+ \pi^- d$ reaction which we discuss in our papers, in particular the $^3D_3$ partial wave, see discussion at the end of Ref.~\cite{tsikeno}. At the energy of the peak of the $pn(I=0) \to \pi^+ \pi^- d$ reaction the  $np \to np$ amplitude will have an enhanced imaginary part due to the optical theorem and this has effects  in the phase shift at this energy. This can be said of most of the reactions claimed to see the dibaryon. What they see is a consequence of the peak seen in the $pn(I=0) \to \pi^+ \pi^- d$ reaction, whatever the reason for this peak be . This is the important point. The peak of the $pn(I=0) \to \pi^+ \pi^- d$ reaction will have repercussion in many observables, but this does not tell us that the reason for the peak has to be a dibaryon. Whatever the reason be, it will have consequences. Actually, the repercussion of the peak of the $pn(I=0)\to \pi^+\pi^-d$ reaction on the $^3D_3$ and $^3G_3$ partial waves was already discussed in \cite{wilkin,miguel}.
  
  The pole or resonant structure is guaranteed by the triangle singularity of the last step $pp\to\pi^+d$ shown in \cite{tsikeno}. It is well-known that a triangle singularity produces an Argand plot similar to the one of a resonance \cite{compass, guofeng}.
\item[vii)] The comment of this point is again weak. It mentions that the ``$d^*(2380)$'' has been seen in the $\gamma d\to d\pi^0\pi^0$ and $\gamma d\to pn$ reactions. Actually in Refs. \cite{Ishikawa1,Ishikawa2} what one observes is a deviation of the experimental cross section from the theoretical calculations of \cite{fix,egorov} at low photon energies. These calculations are based on the impulse approximation and the $\pi$ rescattering terms are neglected. Actually, the rescattering contributions of pions are important, particularly at low energies because the momentum transfer is shared between two nucleons and one picks up smaller deuteron momentum components where the wave function is bigger. Thus, concluding that the discrepancies seen with experiment of a calculation based upon the impulse approximation are due to the dibaryon is an incorrect conclusion.
   
   Actually, we can add more to this discussion. The authors of Refs. \cite{Ishikawa1,Ishikawa2} also studied the $\gamma d\to \pi^0 \eta d$ reaction \cite{ishi1,ishi2}. This reactions was throughly studied theoretically in \cite{alber}, where it was found that the pion rescattering mechanisms were very important, and the most striking feature of the reaction, the shift of the shape of the invariant mass distributions was well reproduced. In the $\gamma d\to \pi^0\eta d$ reaction the $\eta$ rescattering had a small effect, only the $\pi$ rescattering was relevant. In the $\gamma d\to \pi^0\pi^0 d$ reaction the two pions can rescatter making the rescattering mechanism in $\gamma d\to \pi^0\pi^0 d$ even more important than in $\gamma d\to \pi^0\eta d$.
   
   The $\gamma d\to \pi^0\pi^0 d$ reaction has been measured more accurately in \cite{jude}. The same comments can be done concerning this work, since comparison with the data is done with the impulse approximation of Refs. \cite{fix,egorov}. Actually in that paper three dibaryons are claimed. It is not the purpose of this discussion to polemize with these conclusions but we cannot refrain from noting that a fit to the data with a straight line gives a better $\chi^2$ than the one with the three dibaryons.
   
   Concerning the $\gamma d\to p n$ (or $pn\to \gamma d$) signals observed in polarization observables in \cite{ikeda,bashkanov1,bashkanov2}, the following considerations are in order. The cross section for $\gamma d\to pn (pn\to \gamma d)$ has a clear peak due to the $\Delta(1232)$ excitation around $E_\gamma=260$ MeV \cite{rossi,whis}. This reaction is similar to the $pp\to \pi^+d$ reaction studied by us in Ref.~\cite{tsikeno}, which develops a triangle singularity. It is easy to see using the same procedure as in \cite{tsikeno} that the $\gamma d\to pn$ reaction is also driven by the same triangle singularity. In the cross section one does not see any trace of the ``$d^*(2380)$''. However, it is well known that polarization observables are sensitive to small terms of the amplitudes which do not show in integrated cross sections \cite{luisroca}. Thus, the combined reaction $np\to \pi^+\pi^-d\to \gamma d$ provides a contribution to the $np\to \gamma d$ reaction through an intermediate state which has a peak in the ``$d^*(2380)$'' region. As it is the case in \cite{luisroca}, this small amplitude can show up in polarization observables, justifying the observation of \cite{ikeda,bashkanov1,bashkanov2}. Yet, this cannot be seen as a proof of the existence of a dibaryon since it will occur whichever be the reason for the $pn(I=0)\to\pi^+\pi^-d$ peak. 
  \item[viii)] The point 8) is also illustrative. The authors of the comment claim that the cross section for the $pn (I=I) \to \pi^+ \pi^- d$  in our approach should be about $4$ times bigger than the one for $pn(I=0) \to \pi^+ \pi^- d$ , giving hand-waiving arguments, while experimentally it is about $10$ times smaller. Once again, this has a very easy explanation. As we have commented before, the two step process has two amplitudes: $pn (I=0) \to \pi^- pp$ followed by $pp \to \pi^+ d$, and $pn (I=0) \to  \pi^+ nn$ followed by $nn \to \pi^- d$. In the case of equal momenta of the pions the two amplitudes sum and produce an enhancement of the cross section.  On the contrary, for $I=1$ we have: $pn (I=1) \to \pi^- pp$ followed by $pp \to \pi^+ d$, and $pn (I=1) \to \pi^+ nn$ followed by $nn \to  \pi^- d$. But in this case, the two amplitudes cancel exactly. We proved that analytically, but it has to be like that because the two pions are in $I=1$ , which means p-wave and they cannot go together. 
   \item[ix)] Finally, the comment about the Argand plot has an easy answer, and has been mentioned before. Since the last step of our mechanism contains a triangle singularity, this creates a structure very similar to the one of a normal resonance, as discussed in the paper of the COMPASS collaboration \cite{compass}. 
  \end{itemize}
  \section{Acknowledgments}
  Useful comments from M. Bashkanov are acknowledged. The work of N. I. was partly
supported by JSPS
KAKENHI Grant No. JP19K14709. R. M. acknowledges
support from the Contratación de investigadores de
Excelencia de la Generalitat valenciana (GVA) program No.~CIDEGENT/2019/015 and from the spanish
national Grants No.~PID2019-106080GB-C21 and
No.~PID2020-112777GB-I00. This work is partly supported by
Generalitat Valenciana under contract No. PROMETEO/
2020/023. This project has received funding from the
European Unions Horizon 2020 research and innovation
programme under grant agreement No. 824093 for the
STRONG-2020 project.
\bibliography{biblio2}

\end{document}